\documentstyle[aps]{revtex}

\draft
\input epsf

\begin{document}
\title{Controlled Collapse of a Bose-Einstein Condensate}
\author{J.~L. Roberts, N.~R. Claussen, S.~L. Cornish, E.~A. Donley, E.~A. Cornell\cite{byline} and C.~E. Wieman}
\address{JILA,
  National Institute of Standards and Technology and the University of
  Colorado, and the
  Department of Physics, University of Colorado, Boulder, Colorado
  80309-0440}
\date{\today}
\maketitle

\begin{abstract}
The point of instability of a Bose-Einstein Condensate (BEC) due to attractive interactions was studied.  Stable $^{85}$Rb BECs were created and then caused to collapse by slowly changing the atom-atom interaction from repulsive to attractive using a Feshbach resonance.  At a critical value, an abrupt transition was observed in which atoms were ejected from the condensate.  By measuring the onset of this transition as a function of number and attractive interaction strength, we determined the stability condition to be $\frac{N \mid a \mid}{a_{ho}} = 0.461\pm 0.012 \pm 0.054$, slightly lower than the predicted value of 0.574.
 \end{abstract}


The creation of Bose-Einstein condensation in $^{85}$Rb \cite{Cornish2000} allows detailed control of the atom-atom interactions via a Feshbach resonance \cite{Roberts1998,Tiesinga1993a}.  The magnitude and sign of the interaction between atoms can be tuned to any value:  large or small, repulsive or attractive.  The presence of attractive interactions between the atoms has a profound effect on the stability of a Bose-Einstein condensate (BEC), since a large enough attractive interaction will cause the BEC to become unstable and collapse in some way.  In a confining potential, the kinetic energy of the condensate counteracts the attraction and stabilizes the condensate for small enough attractive interactions.  However, as the magnitude of the attractive interaction energy is increased, either by increasing the interaction strength or the number of atoms, the attractive interaction will eventually overwhelm the kinetic energy and make the condensate unstable.  Our ability to tune the magnitude of the attractive interactions allows us to study the onset of this instability and the subsequent behavior in a highly controlled fashion.

In Ref. \cite{Ruprecht1995a}, Ruprecht and co-workers predict when the condensate will become unstable.  The prediction can be cast in the following form:
\begin{equation}
\frac{N_{cr}|a|}{a_{ho}} = k
\end{equation}
where $k$ is a dimensionless constant, {\it a} is the s-wave scattering length that characterizes the strength of the atom-atom interactions, $N_{cr}$ is the maximum number of atoms that will be stable in the condensate, and $a_{ho}$ = $\sqrt{\frac{\hbar}{m \bar{\omega}}}$ is the mean harmonic oscillator length, which characterizes the kinetic energy in the trap ($\bar{\omega}$ is the geometric mean of the trap frequencies).  Several others \cite{othercollapse} also calculate $k$=0.574 using a variety of methods.

Condensates with attractive interactions have been observed in $^{7}$Li \cite{Bradley1997}, and the maximum number has now been observed to be consistent with the predictions in Refs. \cite{Ruprecht1995a} and \cite{othercollapse}.  There are several differences between these $^{7}$Li experiments and our work, however.  In $^{7}$Li, condensates are formed from a rapidly cooled thermal cloud at a negative scattering length whose value remains fixed, and condensate formation kinetics has been a central study \cite{Gerton2000}.  The condensates grow in number until they reach a critical value and then their number decreases.  Following the decrease, the number begins to grow again and the growth and collapse cycle is repeated as atoms are fed into the condensate from the large thermal cloud that is always present.  These cycles are observed to be variable and so a statistical treatment of post-collapse ensembles has been used to examine the $^{7}$Li growth and collapse process \cite{Sackett1999}.

Here, we used the Feshbach resonance in $^{85}$Rb to tune the scattering length by changing the magnetic field.  After forming stable and nearly thermal-atom-free condensates with positive scattering length (repulsive interactions) and a particular number of atoms, we tuned the scattering length to a selected negative value (attractive interactions) in the vicinity of the expected critical value to study the instability.  This allows us to study the onset of instability and the collapse dynamics in a very controlled manner.  From these studies we were able to confirm the functional form of Eq. (1) and measure the stability coefficient $k$.  Also, the transition from stable to unstable condensates was found to be very sharp.  Finally, the number of atoms remaining in the condensate after the collapse event was measured as a function of the initial number.

The first stage in our experiment was to form a stable $^{85}$Rb condensate.  We used a Double Magneto-Optic Trap, or MOT, system \cite{Myatt1996a} to load a Ioffe-Pritchard-type ``baseball'' trap with $3.5 \times 10^{8}$ F=2 m$_{F}$=--2 $^{85}$Rb atoms for evaporative cooling.  Following the somewhat unusual evaporative cooling scheme discussed in Ref. \cite{Cornish2000} we then cooled the trapped atoms until we had a BEC of $\sim$10,000 atoms at a temperature less than 6 nK.  The magnetic field was then ramped very slowly (800 ms) to a value where the scattering length was near zero.

We then ramped the field \cite {trapfreq} linearly in 200 ms to the desired negative scattering length value and waited at that field for 50 ms.  For these small scattering lengths, the condensates were smaller than the 7 $\mu$m resolution limit of our imaging system and we found that errors in the measurement of number were unavoidable when measuring sub-resolution limit condensates. To avoid these errors, after the 50 ms wait we expanded the condensate by changing the magnetic field to achieve a large positive scattering length (76.7 nm).  This caused the condensate to expand outward rapidly due to the mean-field repulsion and become larger than the resolution limit of the imaging system.  After the condensate had expanded sufficiently, we turned off the magnetic trap suddenly and imaged the sample using destructive on-resonance absorption imaging to determine both the number of atoms and the shape of the condensate.  This was repeated for different values of the magnetic field and numbers of condensate atoms.

The major source of difficulty in this experiment was the variation in the number of atoms in the BEC at the end of evaporative cooling.  Several steps were necessary to reduce the variation to under 10\%:  moving all ferromagnetic materials away from the vicinity of the magnetic trap, reducing table and trap coil vibrations, stabilizing the DC magnetic field to better than 6 ppm, and adding another stage of sample preparation.  This additional stage consisted of waiting 2-10 seconds at the $a\sim$ 0 field to allow density-dependent inelastic losses \cite{Roberts2000} to smooth out number variations.

The condensates showed no effect from their excursion into the region of negative $a$ as long as $|a|$ was below a critical value.  Above that value, two dramatic changes in the condensate were observed.  First, there was a substantial decrease in the number of atoms in the condensate \cite{notburst}.  Second, the condensates' ratio of axial to radial widths was variable and usually different from the initial ratio, indicating that the condensates were oscillating in a highly excited state.  We refer to this abrupt change as the condensate ``collapse.''  The transition from stable to unstable is obviously very sharp (see Fig. 1).  We were not able to determine any width to the transition beyond the 4 mG apparent width arising from the shot-to-shot variations in initial condensate number, and we never observed ``partial'' collapses.

The data in Fig. 1 and other sets like it also show the fraction of atoms that remain after the collapse event for an initial number around 6400.  That fraction was 0.58(3).  Changing the starting number to $\sim$2600 atoms gave a fraction remaining of 0.76(3).  As indicated by the standard deviation bars (Fig. 1), the standard deviation in the observed number of atoms after the collapse was larger than the pre-collapse variation.  For the 6400 atom case the initial number had a fractional standard deviation of 7\%, while the fraction remaining varied by 20\%.  For the 2600 atom case, the pre- and post-collapse variations were 9\% and 20\% respectively.

To examine the validity of the theory prediction in Eq. (1), we varied the number of atoms that were initially in the condensate and then determined the minimum magnetic field at which that number of atoms would collapse.  Since the scattering length varies linearly with magnetic field for the negative scattering lengths of interest, Eq. (1) would predict that a plot of  $\frac {1}{N_{cr}}$ vs. the collapse field would be a straight line.  The slope of this line would then determine the stability coefficient $k$.

The presence of drift and random variation in initial number complicated the measurement by making it impossible to know the exact number of atoms that were in the condensate before the ramp to the negative $a$.  When a collapse event is observed, the initial number of atoms in the pre-collapse condensate is only known as well as the shot-to-shot reproducibility of the experiment.  To monitor the initial number, normalization points for which the condensates were prepared, handled, and measured in an identical manner except that they were not ramped to negative $a$, were frequently interspersed with the points with a ramp to negative $a$. Two days' negative $a$ data are shown in Figs. 2(a) and 2(b).  From data sets such as these we are able to measure the stability coefficient $k$ and test whether or not the onset of instability scales with $Na$.  Figure 2(c) shows a summary of several days worth of data taken in the same fashion.  As is evident in Fig. 2, the onset of instability scales well with the product of $Na$.

The all--or--nothing nature of collapse led us to use an unusual approach to determine the stability coefficient $k$ in the presence of variations in the initial number.  Instead of fitting a line to a set of random data, we had to determine a boundary between collapse and non-collapse events.  Data were taken at fixed high number ($\sim$6500) and a selected low number that was varied.  Many points were taken at the critical magnetic fields at which the initial number variation resulted in some condensates collapsing while others remained stable.  This ensures that some of the stable condensate points are just on the verge of collapsing.  The two stable condensate data points closest to the boundary, one from the low-number set and one from the high-number set, were then used.  The magnetic field drifted by tens of mG from day to day due to variations in the ambient field.  To allow day-to-day comparisons, the value of the high number collapse field was used to measure the field drift.  The remaining closest-to-boundary collapse points for various numbers are shown in Fig. 2(c) and are fit to determine a slope.  While this is not the most efficient way to determine the slope, the drift in initial number complicates more sophisticated analyses and this method already gives a statistical error for $k$ that is much smaller than the systematic errors.  The uncertainty in the individual slope determination was estimated by using a standard Monte-Carlo simulation \cite{numerical} with the observed initial number variation as an input parameter.  The uncertainty predicted by the simulation is consistent with the measured day-to-day variation in the slope.  
The simulation also confirms that there was no significant systematic error introduced in this method of analyzing the data.

The average of all of the data gives a slope of 0.00128(3) (atoms gauss)$^{-1}$ for 1/$N_{cr}$ vs. magnetic field.  The change in magnetic field was determined as in Ref. \cite{Roberts1998}, with a negligible uncertainty of 0.5\%.  The number of atoms was determined from the optical depth of the cloud taking into account the intensity and frequency of the probe beam, optical pumping effects, and the small Doppler shift arising from scattering multiple photons.  The intensity and frequency of the probe were systematically varied to check the scaling of these effects, and from these measurements, we estimate the systematic error in determining the number to be $\pm 10\%$.  To convert gauss to scattering length, the scattering length vs. magnetic field was determined using the rethermalization time technique as in Ref. \cite {Roberts1998}, but with better control of density, temperature, and reproducibility.  Rethermalization data were taken at magnetic fields $B$ = 168, 170, and 251 G.  Those rethermalization times were combined with the number calibration of the absorption imaging system and the relation between elastic cross-section and rethermalization time \cite{Demarco1999} to determine the scattering length at those three fields.  These scattering lengths were then fit to the functional form of the Feshbach resonance $a=a_{bg}*(1-\frac{\Delta}{B-B_{pk}})$ using the previously-determined peak ($B_{pk}$) and the width ($\Delta$) to measure the background scattering length $a_{bg}$ to be -19.9 nm \cite{Derevianko1999}.  From this we determined that the change in scattering length vs. magnetic field near 166.05 G is -1.752(46) nm/G, or -33.1 bohr/G.

Combining all of these calibrations with the measurement of the slope of 1/N$_{cr}$ vs. magnetic field gives a value of $k$=0.461$\pm 0.012$ (statistical) $\pm$ 0.054 (systematic).  The uncertainty is dominated by the systematic uncertainty in the determination of the number \cite{number}. This value is two standard deviations lower than the predicted value of 0.574, which would indicate that the condensates were collapsing at a slightly lower value of interaction strength than was predicted.

We investigated the effects of both finite temperature and condensate dynamics on the condensate stability.  There have been predictions for the effect of finite temperature on the condensate stability in $^{7}$Li \cite{Davis1999,Mueller2000} and $^{85}$Rb \cite{Mueller2000}.  Reference \cite{Mueller2000} predicts that the shift in the stable condensate number between zero temperature and the temperature at which we took our data should be less than 1\%.  We deliberately created hot clouds that had roughly equal numbers of condensate and thermal atoms and performed a collapse measurement.  This was roughly a factor of 10 more thermal atoms than were present in the previous measurements.  We could detect no change in the collapse point between the hot and cold condensates.  To search for a dependence on condensate dynamics, we increased the initial spatial size of the condensate by starting with an initial scattering length of +11 nm instead of zero and used ramp times of both 200 ms and 1 ms to magnify any possible effect of dynamics.  The previous data were taken using a magnetic field ramp duration (200 ms) that was more than twice the period of the lowest collective excitation of the condensate, to avoid exciting the condensate prior to the collapse. This condition is true except for very near the point of instability where the lowest collective excitation frequency must go to zero \cite{review}.  For the 200 ms ramp there was no detectable change in the collapse point between ramps starting at $a$=+11 nm and $a$=0 nm.  The 1 ms, initial $a$=+11 nm ramp gave substantial pre-collapse excitation to the condensate but the apparent value of $k$ only decreased by 8\%.

We have observed that condensates are stable with sufficiently small attractive interactions.  Once the interactions are increased beyond a sharply defined value, the condensates collapse and rapidly lose 25-40\% in atom number.  The exact point at which the onset of collapse occurs has been measured and determined to be $\frac {N_{cr}|a|}{a_{ho}} = 0.461\pm 0.012 \pm 0.054$, which is 25(12)\% lower than the predicted value.  The study of the precise nature of the collapse dynamics and the mechanisms by which the atoms are lost will be a future topic of work by this group.

The authors would like to acknowledge the assistance of John Obrecht in developing methods to analyze our data.  One of us (S.~L. Cornish) would like to acknowledge the support of the Lindemann Foundation.  This work was supported by the ONR and NSF.


\begin{figure}
\caption{Transition from stable to unstable condensates.  This figure shows the fraction of atoms remaining as the magnetic field was ramped to higher magnetic fields (i.e., stronger attractive interactions).  The hatched region shown is the expected 4 mG width due to the measured initial number variation.  The initial BEC was kept close to 6400 atoms.  At the field in the hatched region, the data fell into two groups and the individual points (filled triangles) are displayed rather than the average of those points.  The solid bars show the standard deviation of the points to the left of the hatched region (left bar) and those to the right of the hatched region (right bar).}
\end{figure}

\begin{figure}
\caption{Determining the onset of collapse as a function of condensate number and magnetic field.  Figures 2(a) and 2(b) show data sets used to determine the stability coefficient via the boundary that divides the data between collapse ($\circ$) and non-collapse ($\bullet$) events.  In Fig. 2(a), the data were concentrated in two field regions to accurately measure the slope of the boundary line.  In Fig. 2(b), the initial number was varied over a larger range to illustrate the functional form of the onset of instability.  Due to initial number variations, some collapse points appear to be on the wrong side of the stability boundary.  Figure 2(c) shows stable condensate numbers known to be on the verge of instability (see text) as determined from several data sets like those in Figs. 2(a) and 2(b).  The fit to these points provides the value of $k$.}
\end{figure}


\begin{references}
\bibitem{trapfreq} The frequencies of the trap were $17.24 \times 17.47 \times 6.80$ Hz. Changing the bias magnetic field changed these frequencies by less than 0.07\%.
\bibitem{notburst}  When comparing the number of atoms before and after the collapse event, we only measured atoms that remained in a central component.  In other words, we were not including the atoms in the burst described in Ref. [1].
\bibitem{number} A precise independent determination of the scattering length would provide a calibration of number through the rethermalization measurements and therefore lead to a corresponding improvement in the determination of $k$.
\end{references}
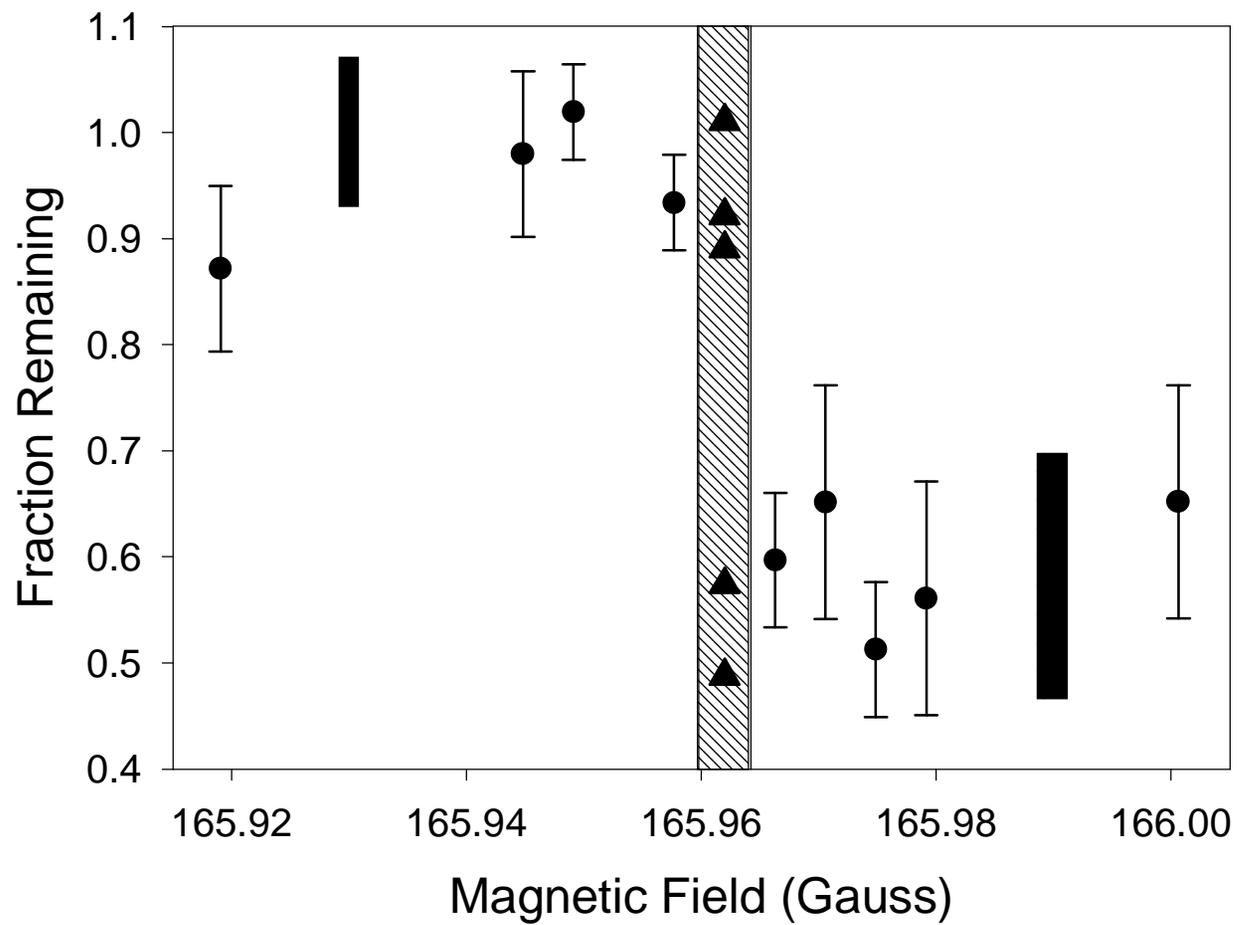
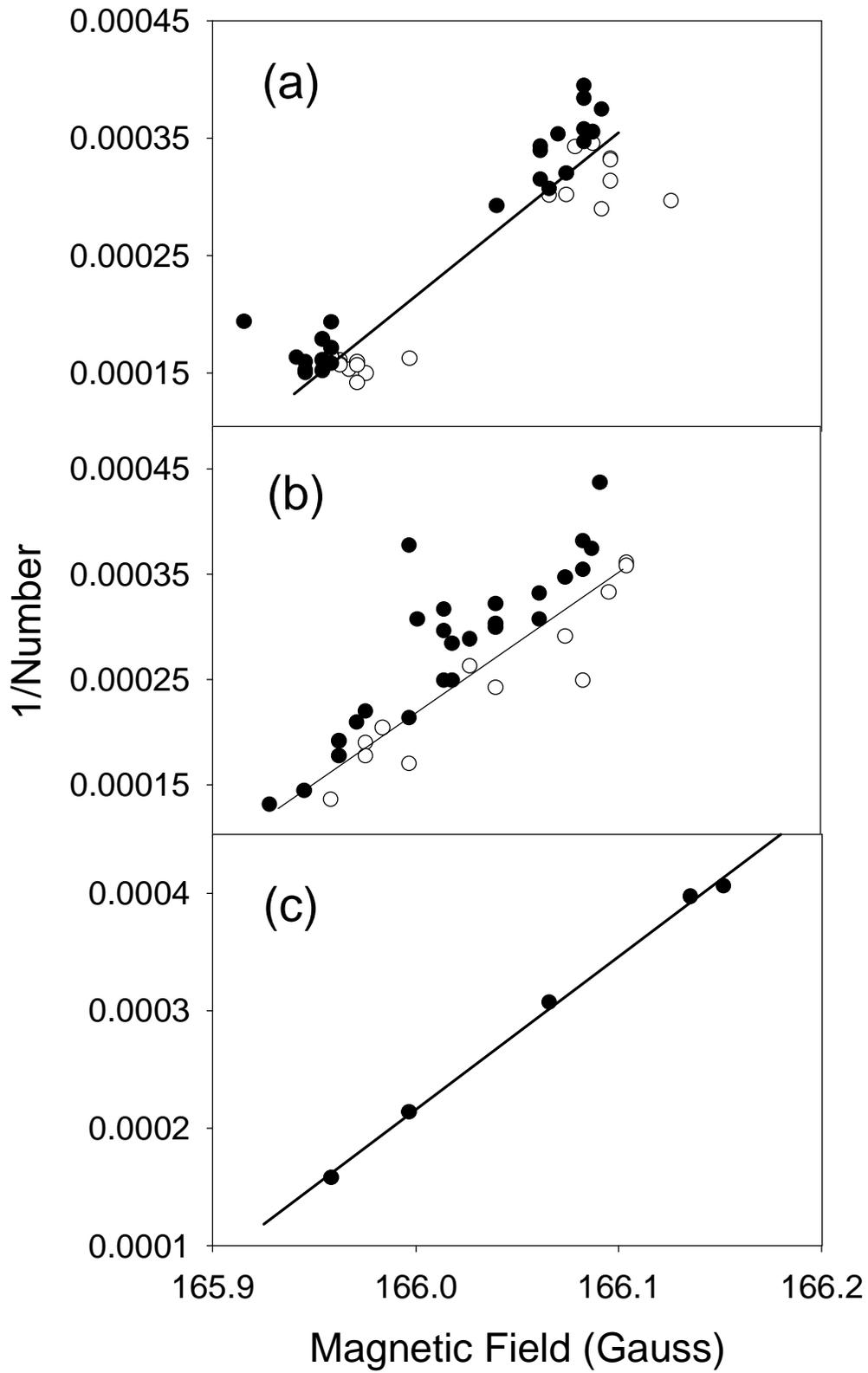
\end{document}